# A new look at weather-related health impacts through functional regression

Pierre Masselot[1*], Fateh Chebana[1], Taha B.M.J. Ouarda[1], Diane Bélanger[1,2], André St-Hilaire[1], Pierre Gosselin[1,2,3]

[1]*Canada Research Chair in Statistical Hydro-Climatology INRS-ETE, Québec, Canada;*

[2]*Centre Hospitalier Universitaire de Québec, Centre de Recherche, Québec, Canada;*

[3]*Institut national de santé publique du Québec (INSPQ), Québec, Canada.*

*\*Corresponding Author: pierre-lucas.masselot@ete.inrs.ca*

June 2018

# Abstract

A major challenge of climate change adaptation is to assess the effect of changing weather on human health. In spite of an increasing literature on the weather-related health subject, many aspect of the relationship are not known, limiting the predictive power of epidemiologic models. The present paper proposes new models to improve the performances of the currently used ones. The proposed models are based on functional data analysis (FDA), a statistical framework dealing with continuous curves instead of scalar time series. The models are applied to the temperature-related cardiovascular mortality issue in Montréal. By making use of the whole information available, the proposed models improve the prediction of cardiovascular mortality according to temperature. In addition, results shed new lights on the relationship by quantifying physiological adaptation effects. These results, not found with classical model, illustrate the potential of FDA approaches.

Climate change adaptation is an important challenge for years to come. With the expected increasing number of heat wave events as well as other meteorological changes, human health could greatly be impacted by climate change[1]. As a consequence, the number of studies outlining the impacts of weather on human health has grown rapidly during the last few years[2–4]. Health issues of interest can either be general mortality or morbidity[5–7], specific diseases such as respiratory illnesses[8,9] or cardiovascular diseases (CVD)[10–13].

In order to estimate the relationship between a meteorological exposure and a health issue, studies commonly use nonlinear models allowing for delayed effects between the exposure and the response. These models can either be generalized additive models (GAM) with lagged exposures[14] or distributed lag nonlinear models (DLNM)[15]. These models have allowed, for instance, the estimation of the acute effect of heat as well as the more latent but very important effect of cold on mortality[3]. However, they are not able to estimate other important aspects of weather-related health, such as the impact of seasonal anomalies (*e.g.* early heat waves or late cold spells) or the potential impact of intra-day variations. Not accounting for these effects significantly hinders the predictive performances of classical models and thus the assessment of the impacts of climate change. As a consequence, there is a need for innovative statistical models.

The present study aims at modelling two little known aspects of temperature-related cardiovascular mortality in the Montréal's metropolitan area (Canada): i) the effect of the previous 24 hour temperatures on mortality during summer and ii) the evolution of the relationship between temperature and mortality along the whole year. To study these aspects, it is hereby proposed to consider the recently developed functional regression models[16]. Functional regression is a subset of functional data analysis models (FDA)[17] which represent a statistical framework to deal with continuous curves (rather called functions or functional data) instead of series of discrete data points. The difference between classical time series and functional data is illustrated in Figure 1. The FDA framework is hereby summoned because of its capacity to represent natural processes such as the intrinsically continuous environmental ones like temperatures. By unveiling the continuous process generating the observed discrete data points, the FDA framework makes use of the whole information available. This leads to a better understanding and *in fine*, to better predictive performances of the phenomena of interest.

The first objective of the present study is to model the effect of the previous 24 hour temperature variations on summer cardiovascular mortality in the Montréal's metropolitan area. For this, we utilise hourly measurements of temperature and daily cardiovascular mortality. Classical models are not able to integrate these data directly since the explanatory and response variables are on different time steps (there are 24 temperature values for each mortality observation). Using each hourly measurement as a different explanatory variable in a multiple regression model would lead to unstable estimates because of the high collinearity between them, as well as the high number of coefficients to estimate which increases the uncertainty[18]. These issues are addressed by the functional linear model for scalar response (SFLM, schematized in Figure 2a) which is a linear model taking one or several functional variables as exposure to predict a scalar response. It could be seen as the most popular of all functional regression models[19–21] since it extends this wide problem of predicting a value according to the temporal evolution of one or several predictors. In the SFLM, the successive hourly measurements of temperature during the previous day can be considered as a single functional observation. This extends the classical model since the whole temperature is used (the whole daily curve) instead of an aggregation of this information (*e.g.* mean or maximum temperature) where valuable information could be lost. Therefore, the first part of the study applies the SFLM with the daily mortality count as response and the previous 24 hour temperature curve as explanatory variable (see the Methods section). This first part of the study is thereafter referred to as 'Application 1'.

The second objective of the study is to model the evolution of the relationship between temperature and cardiovascular mortality over the whole year. Epidemiologic studies usually only model the relationship along the temperature dimension, while it is sometimes suggested that the month of occurrence of temperature events also have its importance on mortality[22]. This gap can be filled with the fully functional linear model (FFLM, schematized in Figure 2b), which is even more general than the SFLM since it expresses a functional response according to at least one functional predictor in the most exhaustive way. The second part of the study ('Application 2') uses the FFLM to predict the annual mortality curve using the same year temperature curve. It accounts for the seasonal changes of the relationship (being different between summer and winter for instance). Thus, the FFLM addresses the issue of seasonal anomalies exposed above. In the general model, each time step $t$ of the response curve could depend on *any* time $s$ of the exposure curve (see Figure 2b), rather than the fixed time $t$ only[17]. Since the annual sequence of mortality and temperature data points are considered as a single curve, the FFLM naturally takes into account their smooth variations[16]. Like the SFLM, the FFLM is especially well suited for time series with potential lag effects and, in this case, it is restricted to the influence of the past time period of the temperature on the time step $t$ of cardiovascular mortality.

The next section presents the result of Applications 1 and 2 to show the effect of previous 24 hour temperature curve and the yearly evolution of the relationship. Their results are then compared to those of the classical models applied to the same issues to show the new insights and predictive improvements brought by functional models. These results are then the basis for the discussion about the advantageous properties of functional regression, both qualitatively and quantitatively.

# Results

The data for each of the two following applications are summarized in Table 1. Details about their collection can be found in the Methods section at the end of the manuscript. The Method section also provides mathematical details about the two functional regression models considered in the study.

**Table 1: Summary of the models and data used in the two applications.**

| Application | Model | CVD mortality | | Temperatures | | Time period $T$ | Data period | $N$ |
| --- | --- | --- | --- | --- | --- | --- | --- | --- |
| | | Type | Observation step | Type | Observation step | | | |
| 1 | SFLM | Scalar | Daily | Functional | Hourly | 24 hour | June-August 2007-2011 | 460 days |
| 2 | FFLM | Functional | Daily | Functional | Daily | 365 days | 1981-2011 | 31 years |

## Application 1: daily mortality – hourly temperatures (with SFLM)

Application 1 is a short-term application using the SFLM in order to detail how cardiovascular mortality is impacted by the evolution of temperature during the previous day. Application 1 focuses on summer mortality (*i.e.* June to August) in order to increase the knowledge about the cardiovascular mortality response to summer heat. This period of the year is especially of major importance because of heat waves and their importance in climate change adaptation[23–25]. This is particularly true in southern Québec[26].

To be more accurate, the response $y_i$ of the model is the cardiovascular mortality count at day $i$, and does not need any preprocessing to be used in the SFLM. However, the exposure is the 24 hour temperature curve of day $i-1$, $x_{i-1}(s)$ ($s \in [0; 24]$). The functional observations $x_{i-1}(s)$ need to be constructed from hourly temperature measurements (see the Data section), through smoothing. We refer to the Methods section for mathematical details about constructing these functional observations. A representative subset of the resulting curves $x_{i-1}(s)$ is displayed in Figure 3, along with the original hourly measurements. It shows that temperature measurements were already smooth, and that the obtained functional data $x_{i-1}(s)$ fit almost perfectly the measurements.

In the SFLM, since the exposure is functional and the response is scalar, the relationship between them (*i.e.* the regression coefficient) takes the form of a function $\beta(s)$ depicting the particular effect of each hour $s$ of $x_{i-1}(s)$ on $y_i$. This functional coefficient $\beta(s)$, estimated on the Montréal community area data, is shown in Figure 4 along with its 95% confidence interval. In the present application, $\beta(s)$ is interpreted as a lag-response relationship with an increased resolution to estimate the intraday lags. It explains why the ends of the curve do not coincide in Figure 4.

Figure 4 shows a positive effect of the previous 24-hour temperatures for almost all day. The curve especially shows high values during the morning (5-12 AM) and the evening (4-10 PM). This result clearly shows the role of hot mornings and evenings in explaining the next day's cardiovascular over-mortality. It suggests that the presence of periods of lower temperatures during the day is important to avoid over-mortality episodes during the day. Note that this phenomenon is usually acknowledged by the presence of minimum temperatures in the definition of heat waves[27]. This is a result that is expected from reality, but which cannot be shown by commonly used regression models.

**Application 2: annual mortality – annual temperatures (with FFLM)**

Application 2 uses the FFLM to extend the relationship between temperature and cardiovascular mortality to a longer scale. The response in the present application is the annual curve of cardiovascular mortality ($y_i(t)$, $t \in [0; 365]$) and the predictor is the curve of temperature during the same year ($x_i(s)$, $s \in [0; 365]$). Therefore, Application 2 focuses more on the overall relationship between temperature and cardiovascular mortality. Its main interest is to transcribe the evolution of the relationship shape over the year, which is expected to be different according to the season.

One could have noted that the response is considered as a continuous function although mortality is discrete by nature. From a mathematical point of view, an underlying Poisson process actually generates death occurrences which are then counted each day to yield the classical mortality data[28]. Available daily data are in fact counts from occurrences happening continually which are generated from a continuous rate function (the expected number of occurrences)[29]. Therefore, constructing functional data by smoothing the daily count data consists in estimating the underlying rate function.

Application 2 uses the FFLM meaning that, for both the cardiovascular mortality response and the temperature exposure, functional observations need to be constructed from daily measurements. The resulting curves for year 1983 are displayed in Figure 5 as a representative example. Note that the smoothness of the resulting curves is objectively determined through a cross-validation criterion (CV, see Methods section). Figure 5a shows that the smoothing of cardiovascular mortality considers high-frequency variations as noise, but that the curves represent the overall evolution over the year. This model complements Application 1 by focusing on the long term evolution of mortality. In contrast to mortality, constructed temperature series (Figure 5b) match almost perfectly the measured data points.

Since both the response and the exposure are functional, the relationship between them takes the form of a bi-dimensional functional coefficient, *i.e.* a surface. In addition, this surface is constrained to be non-null only when $s \in [t - 60; t]$, which means that it represents the relationship between the mortality and the temperature curves for the preceding 60 days only.

The estimation of the surface obtained with Montréal's data is shown in Figure 6. It shows a very smooth and mostly negative surface, indicating increases in mortality for below average temperatures. Hence, overall cold weather induces more deaths than heat in Montréal. This effect of cold is especially strong during the fall season (September to November) as well as during the spring season (end of April to the beginning of June). Note also that the surface is close to zero in winter and shows even a positive relationship between January temperatures and February/March mortality. The latter suggests a mortality displacement effect following January cold spells, which is not usually observed during this period of the year[30].

The results shown here could suggest an adaptation of the population (either 'sociological' or 'physiological')[31] by showing that below average temperatures cause more excess mortality during the fall season when the population is not yet well prepared, than during winter when it is common. This type of effect cannot be seen with classical methods such as DLNM, and has especially never been observed in the province of Quebec[12,32].

## Comparison with classical models

The use of functional regression models seems to unveil aspects of the relationship that were not detected by the models commonly used in the field. We now wonder if this is relevant, *i.e.* if these aspects help predicting yet unobserved mortality. First, Application 1 (based on the SFLM) concerns the relationship between mortality and the previous 24-hour temperature. For a fair comparison, classical models are applied with the same data as the SFLM, *i.e.* a GAM is performed with mortality count of day $i$ as response and temperature with lag 1 as exposure. In order to cover different cases, several models are fit with different temperature variables, *i.e.* mean, maximum and minimum temperatures although no differences in predictive power have been observed[33], as well as the diurnal range[34].

**Table 2: Root mean square errors (RMSE) computed through leave-one-year-out cross-validation.** * indicates models introduced in the present paper. For each application, the lowest MSE value is in bold.

| Application 1 | | | | | Application 2 | |
|---|---|---|---|---|---|---|
| $GAM_{min}$ | $GAM_{mean}$ | $GAM_{max}$ | $GAM_{dr}$ | SFLM* | DLNM | FFLM* |
| 3.96 | 3.92 | 3.95 | 4.01 | **3.66** | 4.45 | **4.31** |

Table 2 shows the prediction mean square error (RMSE) of each classical model as well as the SFLM. These RMSEs are computed through 5-fold blocked-CV, *i.e.* each data year corresponds to a fold. This particular version of CV is necessary in the case of time series data[35]. Results show that the SFLM do have a lower prediction error. Therefore, considering the whole curve of a day's temperature instead of a single value not only brings new information but also improves predictive ability.

Application 2 (based on FFLM) uses classical daily data (see Table 1) and can be compared to the widely used DLNM[15], which is fitted on the exact same data. The DLNM is fitted with the same lags than the FFLM, *i.e.* 60 days. Apart from the maximum lag, the DLNM specifications follow those used by a recent international study[3], *i.e.* with internal B-spline nodes placed at the 10th, 75th and 90th percentiles along the temperature dimension and at equally spaced values on the log scale for the lag dimension.

Table 2 shows the RMSE of the DLNM and FFLM, obtained through 31-fold blocked CV (one fold per year). For the FFLM, the predicted mortality curves are d for each day to obtain a daily predicted mortality time series of the same size than the original data. Using this time series, the standard RMSE formula can be computed. It shows RMSE values of 4.45 for the DLNM and 4.31 for the FFLM, indicating hence a better predictive performance for the functional model. Note that, unlike the DLNM, the FFLM achieves these performances while being linear.

Conceptually, the FFLM includes a "time-of-year" aspect, while the DLNM only considers absolute temperatures and is blind to the period of occurrence of temperatures. This time of year aspect could nonetheless be included in a DLNM through an interaction term with a day-of-year variable[36]. Adding this interaction, the RMSE of the DLNM drops to 4.35, which is still slightly above the FFLM's RMSE. This result is not presented here since it does not correspond to the current practice in the literature and has never been performed with the whole year (only summer season).

## Discussion

The present paper intends to introduce functional regression as an alternative approach to classical models in weather-related health studies. Comparison results show that the predictive performance of functional regression is not vastly superior to those of GAMs and DLNMs. Indeed, the differences in RMSE are small compared to the magnitude of daily cardiovascular mortality in Montréal which presents a mean of 17 death and a standard deviation of 5 deaths.

However, functional regression models present several conceptual and practical advantages over the classical models, as discussed below.

The first strength of the functional regression models proposed in the present paper is the natural addition of a time-of-day or time-of-year aspect. Indeed, the SFLM and FFLM acknowledge that temperature may not have the same influence on mortality according to the time in the period studied. This aspect of models allows shedding new knowledge about the relationship between temperature and mortality, both at the intra-day and yearly levels. Note that the two applications respectively consider daily and annual curves since these are the most obvious and frequently used time periods to consider. Other time periods (*e.g.* a week, a month or several years) can however be considered according to the objective of the study, prior physio-pathological knowledge and the available data.

Barnett et al.[33] tried to identify the better measures of temperature for mortality prediction purposes. Evaluating different possibilities such as mean, minimum, maximum temperature as well as other indices based on temperature, the study concluded that none was absolutely better than the others. One of the most interesting properties of the functional framework is that it makes this question irrelevant since the majority of the measures usually considered are automatically included in a temperature curve. In addition, functional regression models address the issue of serial dependence between successive measurements since curves can usually be considered independent[37].

All the strengths of functional regression models exposed above participate to the main interest of the functional framework: unveiling new aspects of the weather-health relationships. The SFLM indicates that temperature has an influence at particular periods during the day, *i.e.* during morning and evening. This latter result corresponds to an intuitive knowledge, since minimum temperatures are included in heat wave definitions to account for this phenomenon[38], but do not usually appear in regression models' results. Note that this knowledge could be further refined by considering temperature curves spanning several days as exposure.

Classical results of temperature-related mortality (usually obtained using a DLNM[15]) mainly represent a dichotomy between extreme heat and cold. Mechanisms accounting for the evolution of the relationship are starting to be included in DLNMs[36,39]. However, this is usually carried out through interaction terms which are difficult to interpret. The FFLM includes these aspects in a straightforward manner and allows a complete and easy interpretation of results. In particular, the FFLM outlines that a cold spell may not have the same impact whether it happens during fall of winter seasons, *i.e.* it allows observing the population adaptation phenomenon (either physiological or sociological habits). These results are obtained by fitting a single model, and support previous evidence[5], obtained through the fit of multiple models. Finally, note that studies on the extremes are also of interest and should complement studies similar to the present one[40].

The functional regression models applied in the present paper are linear for clarity purposes. However, the true relationship might not be linear. Indeed, even when isolating specific periods such as morning and evening for Application 1, or summer and winter for Application 2, the effect could be present only above certain thresholds. A perspective is then to apply nonlinear functional models to similar applications to the ones performed in the present paper. Such models have recently been developed, with for instance the functional GAM[41] or the recently developed general functional models[42,43].

The main limitation of functional models lies in their gluttony of data. Indeed, considering long time periods (such as in Application 2 here) necessitates a large amount of data to obtain a meaningful set of curves. This limitation is exacerbated for nonlinear functional models, further justifying the use of linear ones in the present study. Note, however, that large databases are increasingly available with health and weather services, meaning that this limitation will no longer be relevant in a few years.

Introducing functional models in environmental epidemiology opens new potential directions, but research in this topic needs to be pursued. Indeed, the applications shown here are examples designed to introduce the reader to functional linear models, but merely touched the full range of possibilities offered by the FDA framework. Hence, these models may prove even more useful with more exhaustive applications including several predictors (*e.g.* atmospheric pollutants, precipitations, population demographic data, etc…) and eventually more data curves.

## Methods

### Data

The data are those of the Montréal's metropolitan area, *i.e.* the extended area of the city including all its suburbs. This is the most densely populated area of the province of Quebec and one of the most populous in Canada. The large population allows for enough daily CVD mortality to obtain a meaningful link between mortality and temperatures. Moreover, the area is small and flat enough to consider the weather to be homogeneous[27].

The considered health data corresponds to the daily number of deaths associated to cardiovascular diseases as the primary cause from 1981 to 2011 ($N = 31$ years and $n = 11322$ days). Cardiovascular diseases include ischaemic heart diseases (I20-I25 in the tenth version of the international classification of diseases, ICD-10), heart failure (I50 in the ICD-10), cerebrovascular diseases and Transient cerebral ischaemic attacks (G45, H34.0, H34.1, I60, I61, I63 and 255 I64 in the ICD-10). Corresponding ICD-9 codes are used for the years they were in use during the period of interest (*i.e.* before 2000). The data were supplied by the *Institut national de santé publique du Québec* of the Quebec province.

Temperature data come from two different sources for the two applications (Table 1). For Application 1, data are provided by the *Ministère du Développement durable, de l'Environnement et de la lutte contre les changements climatiques* [Ministry of Sustainable Development, the Environment and Fight against Climate Change]. These data are available hourly from 2007 to 2015. In Application 1, only the hot summer season is considered (June, July and August) since heat waves are susceptible to happen in southern Quebec during these months[44,45]. The common window with mortality data includes the months of June to August of the years 2007 to 2011, which corresponds to $N_1 = 460$ days.

Since Application 2 uses annual curves, temperature data are provided by Environment and Climate Change Canada in order to possess enough years to perform the FFLM analysis. Data corresponds to daily values from 1981 to 2011, which is the same period as the CVD mortality data described previously. Hence, we have $N_2 = 31$ years of data containing 365 days each (note that 29th of February where removed to keep the same number of measurement each year).

**Functional data analysis**

Functional data analysis (FDA) has been introduced by Ramsay in 1982[46] and more recently popularized[17]. FDA is the subject of an increasing amount of research, both in the theoretical and applied sides. The former have led to the development of the functional counterparts of many traditional statistical techniques such as regression models[47–49], principal component analysis[17], canonical correlation analysis[50], classification[51] and spatial methods[52]. In the applied side, the functional framework has improved the modeling in a variety of domains such as ecology[53], hydrology[54–56], marketing[57], medicine[58–60] as well as atmospheric pollution modelling[61].

## Smoothing functional data

Functional data $y_i(t)$, generated by a functional variable $Y(t)$, are continuous curves which can be seen as infinite dimensional vectors. Indeed a value $y_i(t)$ exists for every $t$ in a continuous interval $T$. Functional observations $y_i(t)$ are constructed by smoothing splines from the observed measurements $y(t_l)$. In the present study, the nodes of the smoothing splines are equally spaced along the interval $T$. To limit the number of subjective choices, the number of nodes (which controls the smoothness of resulting curves $y_i(t)$) is chosen by leave-one-out cross-validation.

## Functional linear model for scalar response

The functional linear model for scalar response (SFLM) used in Application 1 is mathematically expressed as

$$\ln(y_i) = s(i) + \beta_0 + \int_0^{24} x_{i-1}(s)\beta_1(s)ds + \epsilon \qquad (1)$$

where $y_i$ is the total mortality count of day $i$, $s(i)$ is a smooth function component designed to control for long-term trend (*i.e.* a natural spline with a knot every three months), $\beta_0$ the traditional intercept, and $x_{i-1}(s)$ ($s \in [0; 24]$) is the temperature curve of day $i - 1$. The functional coefficient of interest is $\beta_1(s)$ which expresses the effect of each time $s$ of the temperature curve on the mortality count $y_i$. The estimation method of $\beta_1(s)$ (and its confidence intervals, CI) can be found in the technical appendix.

Note that FDA lacks asymptotic theory to compute parametric confidence intervals (CI) for $\hat{\beta}_1(t)$. Therefore, the CIs displayed in Figure 4 are computed through the non-parametric method of wild bootstrap[62].

**Fully functional linear model**

The fully functional linear model (FFLM) of Application 2 expresses the cardiovascular mortality function $y_i(t)$ of year $i$, at time $t \in [0; 365]$ as

$$\ln(y_i(t)) = s(i) + \beta_0(t) + \int_{t-60}^{t} x_i(s)\beta_1(s,t)ds + \epsilon_i(t) \tag{2}$$

where $s(i)$ is a smooth function of the year ($i \in \{1, ..., 31\}$) to control for the overall cardiovascular mortality trend, $\beta_0(t)$ is the functional intercept, $x_i(s)$ is the temperature curve of year $i$ and $\beta_1(s,t)$ is the surface indicating the effect of $x_i(s)$ on $y_i(t)$. In Application 2, the constraint is added that $s \in [t - 60; t]$[43]. The latter constraint indicates that the maximum lag chosen is 60 days, a value selected because it is slightly higher than the maximum lag found in the literature for mean daily temperature, which is 40 days[63]. As for the SFLM, the estimation details are deferred to the technical appendix.

## Acknowledgements

The authors wish to thank the *Fonds vert du gouvernement du Québec* for funding the study, and the *Ministry of Sustainable Development, Environment and the Fight against Climate Change* as well as *Environment and Climate Change Canada* for access to temperature data. The analyses were performed using the packages *fda*[64] and *FDboost*[65] of the R software[66]. Finally, the authors would like to thank Jean-Xavier Giroux and Yohann Chiu of INRS-ETE for their relevant comments throughout the study. The authors thank the Editor, Prof. Andy Morse, and two anonymous reviewers whose comments helped considerably improve the quality of the manuscript.

## Authors contributions

PM performed the study, including the statistical analysis and writing the manuscript. FC gave the guidelines of the study, supervised the statistical analysis helped writing and editing the manuscript. TO helped writing and editing the manuscript. DB and PG gave access to data, directed the epidemiologic problematic and helped writing the manuscript. Finally, AS helped writing the manuscript.

## Additional information

A technical appendix detailing the models' estimation complements the manuscript.

The authors declare no competing interests.

# Figure legends

**Figure 1: Difference between classical data points (a) and functional data (b).**

**Figure 2: Schematic illustration of the functional linear model for scalar response (a) and the fully functional linear model (b).**

**Figure 3: Examples of daily temperature curves $x_{i-1}(s)$ used as predictors in the SFLM along with their measurement points.**

**Figure 4: Functional coefficient estimating the relationship between daily mortality counts and the previous day's temperature curve.** The continuous line is the coefficient itself and the dashed lines indicate its 95% confidence interval estimated through 500 wild bootstrap replications.

**Figure 5: Estimated functional data for year 1983 as a representative example.** The continuous line represents the functional data of mortality (a) and temperature (b) while grey points indicate original data points.

**Figure 6: Estimated relationship between the annual mortality curve and the temperature of the same year.** The color represents the value of the coefficient, red being positive, black negative and white null. Note that the seemingly low values of the coefficient are explained by its continuous nature (the relationship is spread across the whole surface).